\documentclass[runningheads]{llncs}
\usepackage[T1]{fontenc}
\usepackage{graphicx}
\usepackage{amsmath}
\usepackage{multirow}
\usepackage{booktabs}
\usepackage{amssymb}
\usepackage[colorlinks=true,citecolor=blue,urlcolor=blue]{hyperref}
\usepackage[misc]{ifsym} 
\usepackage{subfigure}
\usepackage{color}
\usepackage{algorithm}
\usepackage{algorithmic}
\begin{document}
\title{{Dynamic Position Transformation and Boundary Refinement Network for Left Atrial Segmentation}}  

\titlerunning{DPBNet}

\author{Fangqiang Xu\inst{1}%index{Xu, Fangqiang}
    \and
     Wenxuan Tu\thanks{Corresponding Author.}\inst{2} %index{Tu, Wenxuan}
    \and
    Fan Feng \inst{1} %index{Feng, Fan}
    \and
    Malitha Gunawardhana\inst{1} %index{Gunawardhana, Malitha}
    \and
    \\Jiayuan Yang\inst{1}%index{Yang, Jiayuan}
    \and 
    Yun Gu\inst{1}%index{Gu, Yun}
    \and 
    Jichao Zhao\inst{\star 1}%index{Zhao, Jichao}
 }
 
\authorrunning{F. Xu et al.}

\institute{Auckland Bioengineering Institute, University of Auckland\\
\email{johnnyxu158@gmail.com, j.zhao@auckland.ac.nz}
    \and School of Computer Science and Technology, Hainan University \\
    \email{wenxuantu@163.com}
 }

\maketitle
\setcounter{footnote}{0}

\begin{abstract}
Left atrial (LA) segmentation is a crucial technique for diagnosing irregular heartbeat (i.e., atrial fibrillation). Most current methods for LA segmentation strictly assume that the input data is acquired using object-oriented center cropping, while this assumption may not always hold in practice due to the high cost of manual object annotation. Random cropping is a straightforward data pre-processing approach. However, it 1) introduces significant irregularities and incompleteness in the input data and 2) disrupts the coherence and continuity of object boundary regions. To tackle these issues, we propose a novel \textbf{D}ynamic \textbf{P}osition transformation and \textbf{B}oundary refinement \textbf{Net}work (DPBNet). The core idea is to dynamically adjust the relative position of irregular targets to construct their contextual relationships and prioritize difficult boundary pixels to enhance foreground-background distinction. Specifically, we design a shuffle-then-reorder attention module to adjust the position of disrupted objects in the latent space using dynamic generation ratios, such that the vital dependencies among these random cropping targets could be well captured and preserved. Moreover, to improve the accuracy of boundary localization, we introduce a dual fine-grained boundary loss with scenario-adaptive weights to handle the ambiguity of the dual boundary at a fine-grained level, promoting the clarity and continuity of the obtained results. Extensive experimental results on benchmark datasets have demonstrated that DPBNet consistently outperforms existing state-of-the-art methods.

\end{abstract}

\section{Introduction}
Atrial fibrillation (AF), a prevalent arrhythmia, is primarily treated with left atrial (LA) catheter ablation when pharmacological approaches prove insufficient~\cite{burstein2008atrial,zhao2012image}. Considering the variability in patients' heart anatomies, minor errors in atrial structure analysis can significantly impact disease diagnosis. Therefore, accurate segmentation of the left atrium is both crucial and challenging.

According to the learning paradigm, existing left atrial segmentation methods can be roughly categorized into one-stage and two-stage models~\cite{vesal2019dilated,xia2019automatic,xiong2018fully,xiong2020fully}. Two-stage models ensemble two decoupled networks to generate segmentation results by detecting region of interest (ROI), while one-stage model is known for its simplicity and effectiveness, by directly predicting results. The key prerequisite for the success of current one-stage methods lies in the assumption that all samples have a centrally complete foreground target~\cite{liu2022uncertainty,ronneberger2015u,uslu2021net,wu2021semi}.
However, such an assumption may not always hold in practical scenarios considering the difficulty in collecting definite central data, influenced by diverse equipment and variable atrial structures in patients~\cite{hansen2015atrial,zhao2017three}. For instance, medical experts must spend a substantial amount of time cropping to extract centralized left atrial data from different, which contradicts the goal of developing a fully automated model to reduce human effort and time.

\begin{figure}
\includegraphics[width=\textwidth]{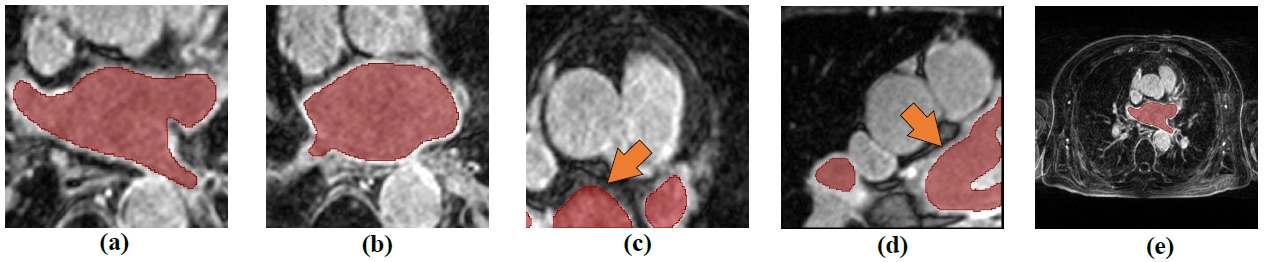}
\caption{Input comparison showcases center cropping in (a), (b), random cropping in (c), (d), with all cropping applied to the original image (e). Red areas represent ground truth, while orange arrows highlight irregular object and discontinuous boundary in random cropping outcomes.
} \label{fig1}
\end{figure}

Randomly cropping a fixed-size input from original data is a straightforward processing strategy~\cite{chen2019multi}. However, it has two significant drawbacks that impact model performance. As seen in Fig.~\ref{fig1}, \textbf{1) Random cropped inputs exhibit greater irregularity and incompleteness compared to central inputs.} This irregularity disrupts original positional relationships, posing challenges for most existing convolutional neural networks (CNNs)~\cite{2023HCHSM,2024AMGC,2024RITR} to accurately capture contextual dependencies. Especially in cases involving irregular small targets with unknown range relationships, existing networks often struggle to capture effective dependencies among these irregular target regions. As a consequence, this leads to decreased model confidence and prediction accuracy.
\textbf{2) Random cropping disrupts the coherence and continuity of boundary regions.} Current loss functions typically treat all pixels equally, neglecting the importance of boundary optimization. Despite advancements in boundary weighting techniques, accurately discerning fine-grained distinctions at dual (foreground-background) boundaries remains challenging. This challenge is particularly prominent when using random cropping, as it can lead to ambiguity or overflow in the boundary regions. In light of this, we find a random-cropped input framework has been an urgent yet under-developed requirement for LA segmentation. 

Based on the aforementioned observations, we propose a novel \textbf{D}ynamic \textbf{P}osition transformation and \textbf{B}oundary refinement \textbf{Net}work (DPBNet) for Left Atrial Segmentation. The key idea is to dynamically adjust the relative position of irregular targets to construct their contextual relationships and focus more on hard boundary pixels to enhance foreground-background distinction.
Specifically, we design a Shuffle-then-Reorder Attention Module (SRAM) that utilizes adaptive shuffle ratios to dynamically adjust the positions of feature maps. By incorporating the shuffle-then-reorder operation, the network is encouraged to effectively identify critical information for irregular and incomplete targets, regardless of positional constraints. Moreover, we introduce a dual fine-grained boundary (DFB) loss to enhance boundary precision by applying scenario-specific weights at boundary regions. Unlike straightforward weighting towards the foreground edge~\cite{uslu2021net,xia2021edge}, the DFB Loss distinguishes the fine-grained optimisation difficulties of dual (fore and background) boundary points, effectively optimizing complex boundary regions even with discontinuity inputs. 
By incorporating these techniques, DPBNet is capable of adapting to various random inputs and achieving a more precise focus on boundary details. This enables the network to generate more reliable intermediate features, ultimately leading to improved performance.

The main contributions of this work are summarized as follows:
\textbf{1)} To the best of our knowledge, this is the first attempt to investigate a random cropping input one-stage framework, which is more practical for Left Atrial Segmentation. \textbf{2)} A novel Shuffle-then-Reorder Attention Module (SRAM) is proposed to obtain dynamic position dependencies for random cropping inputs. Moreover, we introduce a dual fine-grained boundary (DFB) loss to enable more precise optimization of ambiguity regions by assigning different weights to boundary conditions. \textbf{3)} Extensive experimental results have demonstrated that our method can achieve state-of-the-art performance compared to existing methods.

\section{Methods}
\begin{figure}
\includegraphics[width=\textwidth]{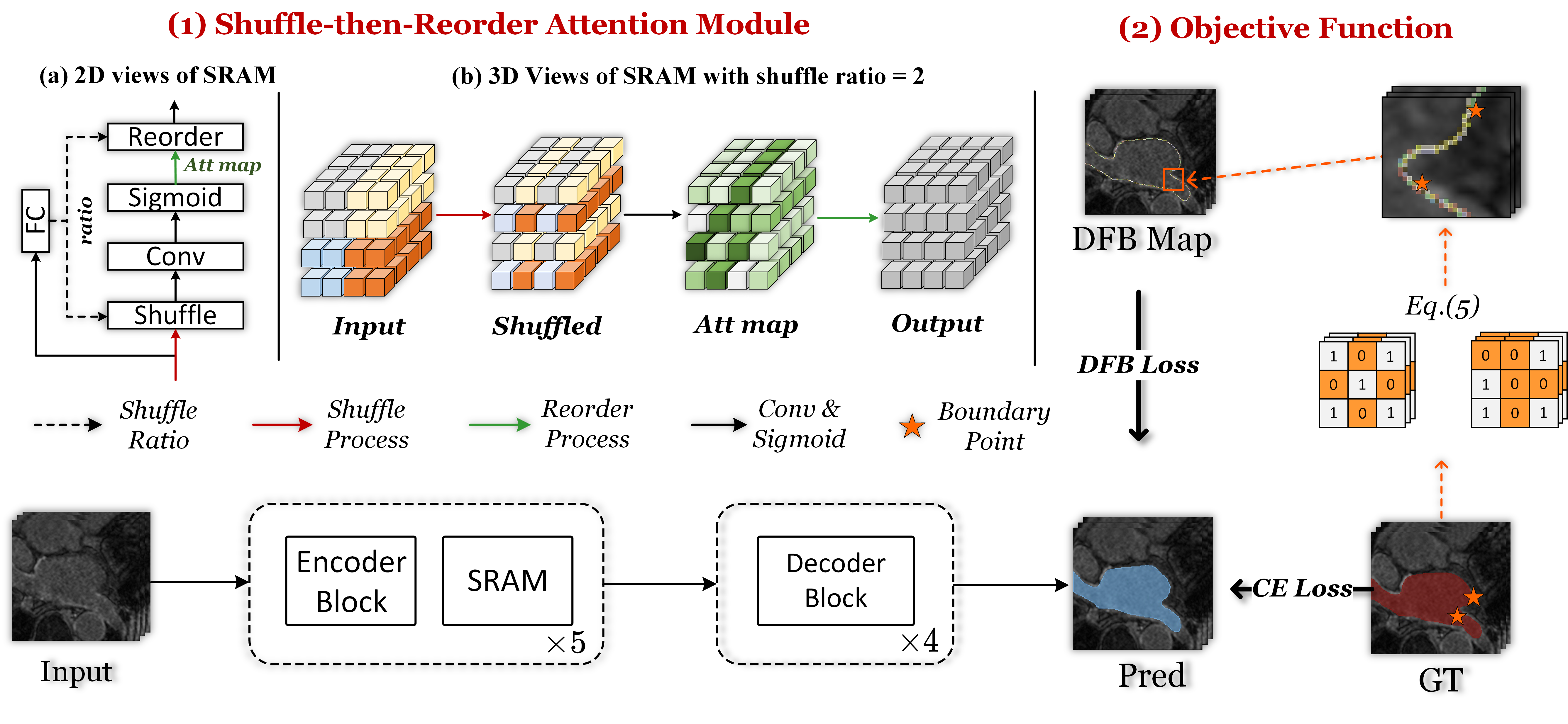}
\caption{The pipeline of our Dynamic Position transformation and Boundary refinement framework utilizes VNet as its backbone architecture. Our learning loss consists of a cross-entropy loss and a dual fine-grained boundary loss on the DFB map.} \label{fig2}
\end{figure}

Based on the aforestated analysis, we are confronted with a challenging yet under-explored task: utilizing irregular and discontinuous input to generate accurate segmentation results. An intuitive solution is to manually modify data sequences (e.g., shifting) and magnify foreground-background boundary differences. To achieve this, we propose a Dynamic Position transformation and Boundary refinement Network (DPBNet) with a Shuffle-then-Reorder Attention Module (SRAM) and the Dual Fine-grained Boundary Loss (DFB Loss), as shown in Fig.~\ref{fig2}. The following sections will provide a detailed explanation of the SRAM, followed by a comprehensive exploration of the DFB Loss function.

\subsection{Shuffle-then-Reorder Attention module}
\subsubsection{Shuffle-then-Reorder Operation.} 
Inspired by the shift operation of the Swin Transformer~\cite{liu2021swin,jiang2022stacked,jiang2019channel}, we devise a shuffle-then-reorder operation with generated shuffle ratios. The key idea is to dynamically adjust the positions of feature maps to capture crucial information that may be easily overlooked due to position separation. As illustrated in Fig.~\ref{fig2}, we define the shuffle operation on ${F \in \mathbb{R}^{C \times H \times W \times D}}$ to generate the shuffled feature map ${F_s \in \mathbb{R}^{C \times H \times W \times D}}$, by adjusting feature map positions, which can be formalized as:
\begin{equation}\label{eq:1}
F_s[:, i^h_s, i^w_s, i^d_s] = F[:, \kappa(i^h), \kappa(i^w), \kappa(i^d)]
\end{equation}
where $i^x$ and $i_s^x$ denote positions before and after shuffling along dimension $x$, while the symbols C, H, W, and D denote the number of channels, height, width, and depth within a feature map, respectively. Consequently, the shuffled indices $i^x_s$ for each dimension $x\in {\{ h,w,d} \}$ are calculated using the following formulation:
\begin{equation}\label{eq:2}
\kappa(i^x)=((i^x-1) \bmod g^x \times r^x) + \left \lfloor \frac{i^x-1 }{g^x} \right \rfloor +1 
\end{equation}
where $\bmod$, $r^x$ and $g^x$ refer to the modulo operation, shuffle ratio and the number of groups in each dimension, respectively, satisfying their product equals the dimension size (e.g., $H=r^h \times g^h$). 
Similarly, the reorder operation is applied inversely to $F_s$, utilizing these shuffled indices to restore $F$ to its original spatial position. This ensures consistency in the feature representations.

It is worth noting that the shuffle and reorder processes can be implemented using a simple matrix transpose operation.
By utilizing these operations, the model dynamically adjusts the positions of irregular objects within the latent space. This allows for establishing appropriate distance relationships for such targets without being confined to predetermined input positions.

\subsubsection{Shuffle-then-Reorder Attention Module.} 
Accordingly, following the shuffle-then-reorder principle, we design a spatial attention-style approach to capture semantically related object dependencies from local and global aspects.
Firstly, we apply max and average pooling operations along the channel and spatial dimensions to generate two feature descriptors, namely $F_{MAP}^c$ and $F_{MAP}^s$. To obtain dynamic shuffle ratios $r^h,r^w$ and $r^d$, we perform a standard linear layer on $F_{MAP}^c$:
\begin{equation}\label{eq:3}
{r_{h}, r_w, r_d} = W(F_{MAP}^c),
\end{equation}
where $W(\cdot)$ and ${\{r^h,r^w,r^d\}}$ refer to linear layer and shuffle ratios, respectively. 
Next, to break the positional constraints of irregular and incomplete objects, we perform shuffle and convolution operations on $F_{MAP}^s$ by the generated shuffle ratios. 
Finally, we employ a reorder operation with the Sigmoid function to obtain an attention map, which is then combined with the original feature $F_{org}$ through element-wise multiplication and a residual connection. This process aims to refine the feature maps $F_{ref}$.
The whole process can be represented as:
\begin{equation}\label{eq:4}
F_{ref} = (\Theta(\Phi(F_{MAP}^s)) + 1) \otimes F_{org} 
\end{equation}
where $\Phi(\cdot)$ denotes CNN-based shuffle operation, while $\Theta(\cdot)$ denotes reorder operation with Sigmoid function, respectively.
In this way, we present a novel plug-and-play attention mechanism that can be seamlessly integrated into existing architectures to effectively capture both short and long-range relationships. Additionally, by incorporating adaptive geometric transformations on the input data, our model mitigates the challenges of fragmentation and lost connectivity among objects caused by random cropping. This adaptation significantly enhances mutual localization between objects, thereby improving overall performance.

\subsection{Dual Fine-grained Boundary Loss} 
To overcome the limitations of boundary discontinuity, we propose a novel Dual Fine-grained Boundary Loss (DFB Loss). The key idea is to assign different weights to the boundaries of foreground and background regions. This allows us to refine the details of both types of boundaries at a fine-grained level.
Initially, we define a point as an interior point if all pixels within its $k$-neighborhood exhibit uniformity. Conversely, we classify a point as a boundary point if there exists any opposite pixel within this range.
Subsequently, the weight of each point is determined based on the number of opposing labels in its $k$-neighborhood, and interior points are uniformly assigned a weight of 1. This is defined as a dual fine-grained boundary map (DFB Map), and can be represented as follows:

\begin{equation}\label{eq:5}
w_i = \begin{cases} 
k^3 - f^3d(g_i) + 1 & \text{if } i=1; \\
f^3d(g_i) + 1 & \text{if } i=0.
\end{cases}
\end{equation}
where $k^3$ represents the maximum possible count of foreground points within the k-nearest neighbors. Moreover, we employ a 3D-CNNs to accurately determine the actual number of foreground points (with a value of 1) within the k-nearest neighbors of a central point. As illustrated in Fig.~\ref{fig2}, when the central point is classified as a foreground point, we determine the count of surrounding background points by subtracting the output of the 3D-CNNs from $k^3$.
Ultimately, the DFB loss can be formulated as follows:

\begin{equation}\label{eq:6}
\ell_{dfb}=1-2\times \frac{\sum_{i}^{N}w_i \cdot p_i \cdot g_i + \epsilon }{\sum_{i}^{N}w_i \cdot p_i +  w_i\cdot g_i + \epsilon} 
\end{equation}
where $N$ represents the total number of points, $p_i$ denotes the predicted probability for point $i$, $g_i$ represents the ground truth for point $i$, and $\epsilon$ is a small constant introduced to avoid division by zero. Consequently, the gradient of DFB Loss can be written as:
\begin{equation}\label{eq:7}
\frac{\partial \ell_{dfb}}{\partial p_i} =-\frac{2Ng_iw_i}{Nw_i(g_i+p_i)} + \frac{Nw_i(2Ng_ip_iw_i)+2\epsilon}{(Nw_i(g_i+p_i)+\epsilon)^2} 
\end{equation}

The gradient of the DFB loss signifies that the model assigns varying priorities to different types of boundaries during training. 
In other words, the model places greater emphasis on optimizing the more challenging boundary points, ensuring that these points undergo distinct refinement and receive higher priority in the training process.
By enhancing the contrast between background and foreground at the boundaries, the proposed model achieves sharper boundary results, effectively mitigating the discontinuity issue caused by random cropping. To ensure smoother optimization, we further integrate the Cross-Entropy Loss (CE Loss) to optimize the model: 
\begin{equation}\label{eq:8}
\ell_{total} =\ell_{ce} + \ell_{dfb}
\end{equation}
Here, $\ell_{dfb}$ represents the DFB Loss, $\ell_{ce}$ denotes the CE Loss, and $\ell_{total}$ represents the final combined loss. 

\section{Experiments and results}
\subsubsection{Dataset.}

To evaluate the effectiveness of the proposed DPBNet, we adopt the most commonly used LA dataset from the 2018 Atrial Segmentation Challenge~\cite{xiong2021global}. Specifically, the LA dataset consists of 100 3D late gadolinium-enhanced magnetic resonance images, fully annotated for the left atrial cavity, with an isotropic resolution of $0.625 \times 0.625 \times 0.625$ mm. Additionally, the LA dataset presents images in two distinct resolutions, namely $576 \times 576 \times 88$ and $640 \times 640 \times 88$. In our DPBNet, we maintain the original resolution of the images throughout both the training and testing phases. Subsequently, the dataset was split into 80 training and 20 testing scans, with a 5-fold cross-validation approach for evaluation.

\subsubsection{Implementation Details.}
For training the networks, we employ the Adam optimizer, with a momentum of 0.9 and a weight decay of 1e-4, aligning with common practices. The input images of dimensions $256 \times 256 \times 80$ are obtained through random cropping from the original image data. The training process is conducted in a total of 15000 iterations, initiated with a learning rate of 1e-4.
To evaluate and compare the performance of DPBNet against other methods, we employ four widely utilized metrics: Dice score, Jaccard index, 95\% Hausdorff Distance (HD95), and Average Symmetric Surface Distance (ASSD). 
For tests, a sliding window ($256 \times 256 \times 80$, stride: $160 \times 160 \times 4$) extracts patches from original inputs, with results averaged for final predictions~\cite{li2020shape,yu2019uncertainty,wu2021semi}. All experiments are conducted by PyTorch 2.1 (Python 3.8) on a single NVIDIA Tesla A100 GPU and an Intel Xeon Platinum 8168 CPU @ 2.70GHz.

\subsection{Comparison Results} 
\begin{figure}
\includegraphics[width=\textwidth]{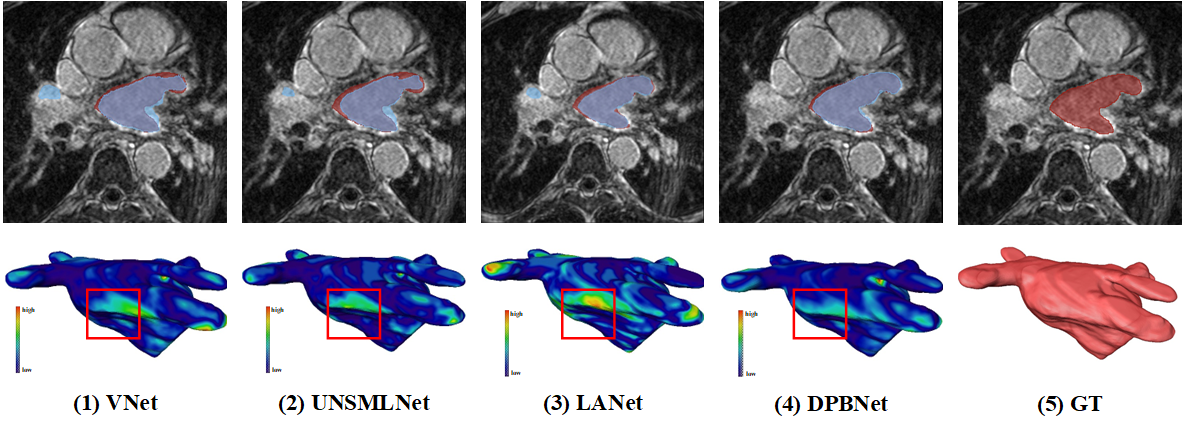}
\caption{
Visualization results by various methods and DPBNet, showcasing prediction (blue) and ground truth (red). First row shows 2D prediction-ground truth overlap, and second features 3D distance maps from prediction-ground truth differences.} \label{fig3}
\end{figure}

To validate the effectiveness of the proposed method, we conduct several experiments with seven state-of-the-art methods on the LA database, which is shown in Table~\ref{tab1} and Fig.~\ref{fig3}. It can be seen from Fig.~\ref{fig3}. that our model generates more accurate and clear predictions compared to state-of-the-art approaches, with DPBNet displaying enhanced boundary details detection as evidenced by yellow areas. Moreover, DPBNet achieves state-of-the-art (SOTA) performance without dependence on any pre-processing or post-processing, directly mitigating challenges from random cropping.  

\begin{table}[h]
\caption{
The comparison across seven existing methods on the LA database highlights top results in bold. Notably, 'C-C', 'R-R' denotes center or random cropping for both input and output, while 'R-C' represents random input with center output. R-DPBNet and C-DPBNet denote testing with random or center data, using same parameters.
}\label{tab1} 
\begin{tabular}{l|c|llll}
\hline
\multicolumn{1}{c|}{\multirow{2}{*}{Methods}} & \multirow{2}{*}{Setting} & \multicolumn{4}{c}{Metrics}                           \\ \cline{3-6} 
\multicolumn{1}{c|}{}                         &                                 & Dice(\%) & Jaccard(\%) & HD95(mm) & ASSD(mm) \\ \hline
VNet~\cite{milletari2016v} (MICCAI'16)                             & C-C                     & 91.85$_{\pm{0.31}}$     & 85.01$_{\pm{0.54}}$        & 2.92$_{\pm{0.32}}$         & 0.96$_{\pm{0.11}}$         \\
Yang et al.~\cite{yang2019combating} (MICCAI'18)                                & C-C                     & 92.24     & 85.64        & -         & 1.49        \\
SEGANet~\cite{lourencco2021left} (MICCAI'20)                           & C-C                     & 91.0$_{\pm{0.2}}$      & 84.0$_{\pm{0.3}}$        & -        & 1.00$_{\pm{0.21}}$         \\
Zhao et al.~\cite{zhao2021not} (ICPR'21)                               & C-C                    & 91.79$_{\pm{1.06}}$      & -            & 2.87$_{\pm{0.67}}$         & -           \\
LANet~\cite{uslu2021net} (TMI'22)                                & C-C                     & 92.0$_{\pm{0.2}}$     & 86.0$_{\pm{0.3}}$         & 2.88$_{\pm{0.56}}$         & 0.86$_{\pm{0.24}}$         \\
UMSMLNet~\cite{liu2022uncertainty} (MP'22)                              & C-C                     & 92.02$_{\pm{0.29}}$        & 85.28$_{\pm{0.49}}$           & 2.84$_{\pm{0.15}}$         & 0.89$_{\pm{0.02}}$        \\ 
CANet~\cite{zhao2023context} (ESA'23)                                & C-C                     & 91.24    & 83.96        & 5.71         & 1.57        \\

\textbf{R-DPBNet (Ours)}                                 & \textbf{R-R}                            & 92.50$_{\pm{0.25}}$      & 86.10$_{\pm{0.44}}$        & 2.83$_{\pm{0.16}}$          & 0.94$_{\pm{0.07}}$        \\

\textbf{C-DPBNet (Ours)}                                 & \textbf{R-C}                            & \textbf{92.57$_{\pm{0.33}}$ }     & \textbf{86.24$_{\pm{0.57}}$ }       & \textbf{2.74$_{\pm{0.27}}$ }         & \textbf{0.85$_{\pm{0.04}}$ }        \\
 \hline
\end{tabular}
\end{table}

Quantitative analysis in Table~\ref{tab1} demonstrates that our R-DPBNet, utilizing random cropping, surpasses competing methods on most metrics, notably achieving a Dice score of 92.50\%. Advantages in HD95 and ASSD for our R-DPBNet are limited by its comprehensive analysis ($640 \times 640 \times 88$) through random cropping, as opposed to the narrower center cropping ($256 \times 256 \times 88$) approach. After employing identical center cropping configurations, as in C-DPBNet, the proposed method in this paper easily achieves the best performance across all metrics. Summarily, our approach breaks impractical assumptions and addresses random cropping challenges, setting new benchmarks for excellence.

\begin{table}[h]
\centering
\caption{Ablation studies of our proposed DPBNet on the LA database.}\label{tab2}
\begin{tabular}{c|c|cccc}
\hline
\multirow{2}{*}{Methods} & \multirow{2}{*}{Setting} & \multicolumn{4}{c}{Metrics}                           \\ \cline{3-6} 
                         &                                 & Dice(\%$\uparrow$) & Jaccard(\%$\uparrow$) & HD95(mm$\downarrow$) & ASSD(mm$\downarrow$) \\ \hline
Vanilla VNet~\cite{milletari2016v}             & R-R                            & 90.77     & 83.37        & 4.94         & 1.37        \\
+ SRAM (k=3)             & R-R                            & 91.81     & 84.95        & 2.96         & 1.44        \\
+ SRAM (k=5)             & R-R                            & 92.21     & 85.61        & 3.11         & 1.06        \\
+ SRAM (k=7)             & R-R                            & 92.16     & 85.54        & 6.83         & 1.25        \\
+ Edge Loss~\cite{uslu2021net}             & R-R                            & 91.67     & 84.72        & 4.60         & 1.46        \\
+ DFB Loss               & R-R                            & 92.29     & 85.74        & 3.89         & \textbf{0.90}        \\
\textbf{+ All (k=5)}              & R-R                            & \textbf{92.33}     & \textbf{85.83}        & \textbf{2.74}         & 1.00        \\ \hline
\end{tabular}
\end{table}

\subsection{Ablation Study}
To validate the contribution of the SRAM and DFB Loss, we conduct an ablation study on the LA dataset, with united framework VNet~\cite{milletari2016v} as the backbone. As shown in Table~\ref{tab2}, all the setting of SRAM generates better results than the baseline, while kernel size 5 gains better performance than other settings. This implies that dynamic adjustments in positioning can enhance the capability of mutual target localization, generating better performance. Moreover, DFB Loss shows significant improvements over both the baseline and Edge Loss, which adopts single and uniform boundary weights. This improvement stems from DFB Loss enhancing boundary refinement and foreground-background contrast.

\section{Conclusion}
In this paper, we address the impractical assumption of current LA segmentation methods on object-oriented center cropping due to high manual annotation costs. Specifically, we propose a new Dynamic Position transformation and Boundary refinement Network (DPBNet) by designing a shuffle-then-reorder attention module and a dual fine-grained boundary loss. The core strategy dynamically enhances object interactions and refines foreground-background details. Comprehensive experiments demonstrate that the proposed DPBNet achieves SOTA results on the LA dataset, to the best of our knowledge.

\begin{credits}
\subsubsection{\ackname} This work was supported in part by the Health Research Council of New Zealand (\#21\/355), the Postgraduate Research Student Support (PReSS) funding in University of Auckland, and Postgraduate Scientific Research Innovation Project in Hunan Province under Grant CX20220076.

\subsubsection{\discintname}
The authors have no competing interests to declare that are
relevant to the content of this article.
\end{credits}

\bibliographystyle{splncs04}
\bibliography{Paper-0950}
\end{document}